\documentclass[12pt]{iopart}
\usepackage{graphicx}
\usepackage{amsbsy}
\usepackage{natbib}

\usepackage{color}
\usepackage{txfonts}
\usepackage{url}
\usepackage[normalem]{ulem}

\DeclareMathSymbol{\varOmega}{\mathord}{letters}{"0A}
\DeclareMathSymbol{\varSigma}{\mathord}{letters}{"06}
\DeclareMathSymbol{\varPsi}{\mathord}{letters}{"09}

\def\aj{AJ}%
\def\araa{ARA\&A}%
\def\apj{ApJ}%
\def\apjl{ApJ}%
\def\apjs{ApJS}%
\def\aap{A\&A}%
\def\icarus{Icarus}%
\def\mnras{MNRAS}%
\def\ssr{Space~Sci.~Rev.}%
\def\nat{Nature}%
\def\gca{Geochim.~Cosmochim.~Acta}%
\def\grl{Geophys.~Res.~Lett.}%
\def\planss{Planet.~Space~Sci.}%
\newcommand {\Msun} {\mbox{M$_{\odot}$}}

\begin{document}

\title[The formation of the solar system]{The formation of the solar system}

\author{ S Pfalzner$^1$, M B Davies$^2$, M Gounelle$^{3, 4}$, A Johansen$^2$, C M\"unker$^5$, \\P Lacerda$^6$,  S Portegies Zwart$^7$, L Testi$^{8, 9}$, M Trieloff$^{10}$ \&  D Veras$^{11}$}

\address{$^1$ Max-Planck Institut f\"ur Radioastronomie, Auf dem H\"ugel 69, 53121 Bonn, Germany}
\address{$^2$ Lund Observatory, Department of Astronomy and Theoretical Physics, Box 43, 22100 Lund, Sweden}
\address{$^3$ IMPMC, Mus\'eum National d'Histoire Naturelle, 
Sorbonne Universit\'es, CNRS, UPMC \& IRD, 57 rue Cuvier, 75005 Paris, France}
 \address{$^4$ Institut Universitaire de France, 103 boulevard Saint-Michel, 75005 Paris, France}
\address{$^5$ Institut f\"ur Geologie und Mineralogie, Universit\"at zu K\"oln, Z\"ulpicherstr. 49b
50674 K\"oln, Germany}
\address{$^6$ Max-Planck-Institut f\"ur Sonnensystemforschung, Justus-von-Liebig-Weg 3, 37077 G\"ottingen, Germany}
\address{$^7$ Leiden University, Sterrewacht Leiden, PO-Box 9513, 2300 RA Leiden, the Netherlands}
\address{$^8$ ESO, Karl Schwarzschild str. 2, D-85748 Garching, Germany}
\address{$^9$ INAF-Osservatorio Astrofisico di Arcetri, Largo E. Fermi 5, I-50125, Firenze, Italy}
\address{$^{10}$ Institut f\"ur Geowissenschaften, Im Neuenheimer Feld 234-236, 69120 Heidelberg, Germany}
\address{$^{11}$ Department of Physics, University of Warwick, Coventry CV4 7AL, United Kingdom}

\ead{spfalzner@mpifr.de}
\vspace{10pt}
\begin{indented}
\item[]November 2014
\end{indented}

\begin{abstract}
The solar system started to  form about 4.56 Gyr ago and despite the long intervening time span, there still exist several clues about its formation. The three major sources for this information are meteorites, the present solar system structure and  the planet-forming systems around young stars. In this introduction we give an overview of the current understanding of the solar system formation from all these different research fields. This  includes the question of the lifetime of the solar protoplanetary disc, the different stages of planet formation, their duration, and their relative importance. We consider whether meteorite evidence and observations of protoplanetary discs point in the same direction. This will tell us whether our solar system had a typical formation history or an exceptional one. There are also many indications that the solar system formed as part of a star cluster. Here we examine the types of cluster the Sun could have formed in, especially whether its stellar density was at any stage high enough to influence the properties of today's solar system. The likelihood of identifying siblings of the Sun is discussed. Finally, the possible dynamical evolution of the solar system since its formation and its future are considered. 

\end{abstract}

%
%
%
%
%
\section{Introduction}
For decades the solar system was assumed to be the prototype for planetary system formation. With the detection of over a thousand confirmed exoplanets and 
many more candidates, it has become apparent that many planetary systems exist that differ substantially in their structural properties from our solar system. Nevertheless the formation of the solar system is still of special interest for several reasons. First, it is only for the solar system that we can directly examine material that is left over from the formation process in the form of meteorites. Second, only for the solar system do we have detailed structural information about the entire system including its smaller bodies. Last but not least, it is only
for the solar system that we know for sure that life exists. 

The three major sources about the formation of the solar system are meteorites, the present solar system structure and  contemporary young planet-forming systems. We start by reviewing the current status of meteorite research concerning the chronology of early solar system formation  including the formation of the terrestrial planets in section 2. In this context the question of the origin of short-lived radioactive nuclei in these meteorites is of special interest. Some of these can only be produced in supernovae events of high-mass stars - different possibilities are discussed in section 3. 

Other sources of information are young stars surrounded by accretion discs from which planetary systems might form. In section 4 the properties of these discs - masses, gas content and chemical composition - are discussed. Estimates of the life times of these discs are given and the consequences for planet formation scenarios are discussed. Section 5 provides a closer look at the different stages of planet formation. Starting from dust grains, then considering pebble-sized objects to planetismals the current state of research is presented.   This is followed by the final step in which planets form. 

Many of these young systems  are part of a cluster of stars. There are several indications that our own solar system also formed as part of a star cluster. Section 6 gives the arguments for such an early cluster environment and discusses the possibilities of finding today stars that formed in the same cluster as our Sun did. Not only the location and masses of the planets but also those of the asteroid and Kuiper belt are characteristics of our solar system that might potentially give clues to its formation. In section 7 the early dynamical evolution of the Kuiper belt is illustrated. Possible scenarios for the late heavy bombardment between 4.0 and 3.7 Gyr ago are discussed. It is still an open question to what degree the solar system characteristics changed since its formation and how stable the solar system is in the long-run. The likely long-term evolution of the solar and other planetary systems is discussed in section 8. This is followed by a summary in section 9. 

First, we look at the information that meteorites give about the formation of the solar system. In order to do so a relative age dating of these meteorites is necessary.

\section{The meteorite record of early solar system evolution}

\subsection{The significance of meteorites}
Studying meteorites from our solar system is the only way to directly constrain timescales of its protoplanetary disc evolution. Most meteorites are older than 4.5 billion years and originate from the asteroid belt. The asteroid belt between Mars and Jupiter provides the only vestige of the planetesimals which were the first larger objects in the protoplanetary disc that provided the building materials for the larger planets. Simulations indicate that it is most likely that the larger planets formed via collisions of such first generation planetesimals (e.g. Wetherill 1990, Chambers 2003). The different groups of meteorites sample these first generation planetesimals and cover the different evolutionary steps of early solar system evolution in great detail. In general, three major groups of meteorites can be distinguished. Chondrites represent unprocessed, brecciated early solar system matter, whereas differentiated meteorites such as achondrites and iron meteorites originate from asteroids that have undergone melting and internal differentiation. These asteroidal melting events were triggered by either decay of short-lived $^{26}$Al or by impact events. Due to the short half life of $^{26}$Al (0.7 Myr), the first heating mechanism is confined to the first 5 million years of solar system evolution.

\subsection{Chondrites and their components}
The oldest dated solar system matter are Ca, Al-rich inclusions (CAIs) in chondritic meteorites that have been dated by the U-Pb method to 4.567-4.568 billion years (Amelin et al. 2002, 2011; Bouvier et al. 2007). CAIs are an important anchor point to constrain the abundance of significant short-lived nuclides such as $^{26}$Al or $^{182}$Hf at the beginning of the solar system. In addition to the long lived U-Pb chronometer, short-lived nuclides with their half-lifes of less than 100 million years enable dating of meteorites and their components at an age resolution as low as several tens of thousands of years. Based on combined U-Pb and Al-Mg chronometry, the ages of chondrules, a major component of chondrites, has been constrained to as late as up to 4 million years after solar system formation (e.g. Bizzarro et al. 2004; Villeneuve et al. 2009). It is currently contentious, as to whether there is a circa 1.5 million years age gap between the formation of the first CAIs and the formation of the first chondrules (see Villneuve et al. 2009; Amelin et al. 2011; Larsen et al 2011; Connelly et al. 2012). There is, however, now consensus that the undifferentiated asteroidal parent bodies of chondrites themselves accreted ca. 2-4 million years after the beginning of the solar system (e.g. Bizzarro et al. 2005; Kleine et al. 2008). Because of their younger accretion ages, chondrites escape internal heating triggered by $^{26}$Al (Trieloff et al. 2003, Henke et al. 2013), and thus they preserved their pristine structure.

\subsection{Differentiated asteroids}
Recent high precision dating using Hf-W and Al-Mg chronometry has shown that some differentiated meteorite groups do in fact predate formation of chondrites. Beside CAIs, some groups of magmatic iron meteorites that represent the cores of differentiated asteroids have been shown to constitute the oldest solar system objects (Kleine et al. 2005). If corrected for cosmic ray exposure, magmatic iron meteorites can be shown to have differentiated within the first 2 million years of solar system history, and their asteroidal parent bodies must have accreted earlier than 0.5 million years after solar system formation (Kruijer et al. 2013, 2014; Wittig et al. 2013). Some older groups of achondrites that sample the outer silicate portion of differentiated asteroids differentiated only slightly later than iron meteorites. The particularly pristine group of angrites (achondrites consisting mostly of the mineral augite) is inferred to have differentiated by around 3-5 million years after the CAIs (Larsen et al. 2011, Brennecka \& Wadhwa 2012; Kleine et al. 2012). Likewise, achondrite groups such as acapulcuites, eucrites, and winonaites as well as non-magmatic iron meteorites differentiated in the time interval between 3-5 million years after solar system formations, with inferred accretion ages between 1 and 2 million years after solar system formation (Schulz et al. 2009, 2010). 

\subsection{The new early solar system chronology}
In summary, there is now clear evidence from short-lived nuclide and long-lived U-Pb chronometry that CAIs are the oldest solar system objects, followed by magmatic iron meteorites (differentiation of 1-2 million years after solar system formation) and most achondrite groups and non-magmatic iron meteorites (differentiation 3-5 million years after solar system formation). Differentiation of these asteroidal bodies was likely driven by internal heat sources, with decay of $^{26}$Al being the most important one. The parent asteroids of chondrites accreted much later (2-4 million years after solar system formation) than those of the differentiated asteroids (less than 2 million years after solar system formation). It is therefore likely that the chondritic parent bodies did not undergo internal differentiation, because the heat supply from short-lived nuclides such as $^{26}$Al was insufficient, or, alternatively, their parent body size was too small.  Alternatively, undifferentiated chondrites may sample the outer layers of planetesimals that differentiated in their interiors.

\subsection{Age of the inner terrestrial planets}
The larger rocky planets in the inner solar system formed much later than the small asteroids, at timescales of tens of millions of years. Hf-W systematics in terrestrial samples and chondrites constrain the age for the Earth's growth to at least 38 million years (Kleine et al. 2002, Yin et al. 2002; König et al. 2011) and probably as late as 120 million years after solar system formation (Allegre et al. 2008; Rudge et al. 2010). There are fewer indicators to constrain the age of Mars, but estimates reach from 2 to 10 million years after solar system formation (Nimmo \& Kleine 2007; Dauphas \& Pourmand 2011). The uncertainties of these estimates originate from the parameters that are used to model planetary growth and internal differentiation into metal cores and silicate mantles. The small size of Mars, for instance, has been explained by radial migration of Jupiter, suggesting that Mars represents a “starved” planet that could not accrete further because of interaction with Jupiter (Walsh et al. 2011). 

For constraining the age of the Earth, the Moon-forming giant impact event is of great importance, and physical models explaining the formation of the Moon have dramatically changed over the past years (e.g. Canup \& Asphaug 2001; Cuk \& Stewart 2012; Canup 2012). Likewise, the Hf-W and U-Pb chronometry of Earth's formation has undergone significant reinterpretation. Most earlier studies agreed that both the Hf-W and U-Pb chronometers date segregation of the Earth's core during the final stages of Earth's accretion (e.g. Allegre et al. 2008; Rudge et al. 2010, Wood \& Halliday 2005). A combination of experimental and geochemical studies now rather argue that volatile elements such as Pb were delivered late to the growing Earth, implying that the U-Pb systematics of the Earth's silicate mantle bear no chronological significance (Albarede 2009; Ballhaus et al. 2013; Marty et al. 2012; Sch\"onb\"achler et al. 2010).

The geochemical and isotopic inventory of meteorites provides a unique chronological record of early solar system evolution. It took only less than one million years after condensation of the first solid matter to form the first generation of planetesimals. These planetesimals underwent internal heating between 1 and 4 million years after solar system formation, largely caused by $^{26}$Al decay, triggering their differentiation into metal cores and silicate mantles. The parent asteroids of undifferentiated chondritic meteorites formed 1-2 million years later, and therefore these bodies did not differentiate. Formation of the larger terrestrial planets in the inner solar system occurred via collision of smaller asteroidal bodies. The timescales involved in planetary growth are in the order of tens of million years, with Mars being a possible exception due to gravitational interaction with nearby Jupiter. There is growing evidence, that the volatile element inventory of the inner solar system was added late, i.e. during the last stages of planetary growth.

\section{The origin of short-lived radionuclides}

Chondrites have not endured any geological activity since they formed during the first million years of our solar system.  As such, they record physical processes in the solar protoplanetary disc (SPD) but also of the molecular cloud stage, which preceded the formation of the proto-Sun \citep{Wadhwa2007}. 

Chondrites are made of CAIs, chondrules and iron-nickel metal. All these components are cemented by a fine-grained matrix (Fig. 1). A key property of chondrites is the past presence within their components of short-lived radionculides (SLRs). SLRs are radioactive elements, which have half-lives, $T_{1/2}$, smaller than 200 Myr and were present in the early solar system  \citep{Russell2001}. Though they have now decayed, their past presence in chondrites and other meteorites is demonstrated by excess of their daughter isotopes. 

\subsection{ Astrophysical context}
Since 1962 and the discovery of $^{129}$I ($T_{1/2}$ = 15.7 Myr), a long list of SLRs has been identified in meteorites (see the Table of \citet{Dauphas2011}). SLRs are important because they can potentially help to build an early solar system chronology and are the most likely heat source of the building blocks of planets, planetesimals \citep{Wadhwa2007}. In addition, they can help decipher the astrophysical context of our solar system formation. In other words, assuming the Sun was born as most stars in a (Giant) Molecular Cloud, elucidating SLRs' origin is the only tool to trace back what was the relationship of the Sun with other stars born in the same molecular cloud.

The presence in the early solar system of SLRs having the longest half-lives is not surprising. Similarly to stable isotopes they have been produced in the interior of stars, which have preceded our Sun in the Galactic history and were delivered to the Interstellar Medium (ISM) by supernova (SN) explosions and massive star winds. Their abundance is roughly in line with Galactic evolution models \citep{Meyer2000}. The presence of SLRs with shorter half-lives ($T_{1/2}$ $\lesssim$ 10 Myr) is more difficult to understand because their half-life is comparable or smaller than the typical timescales of star formation processes \citep{Williams2010}. Obviously, constraining the origin of SLRs depends also on their initial abundance, which is not always clearly known.

\begin{figure}
\begin{centering}
\includegraphics[width=0.38\textwidth, angle =-90]{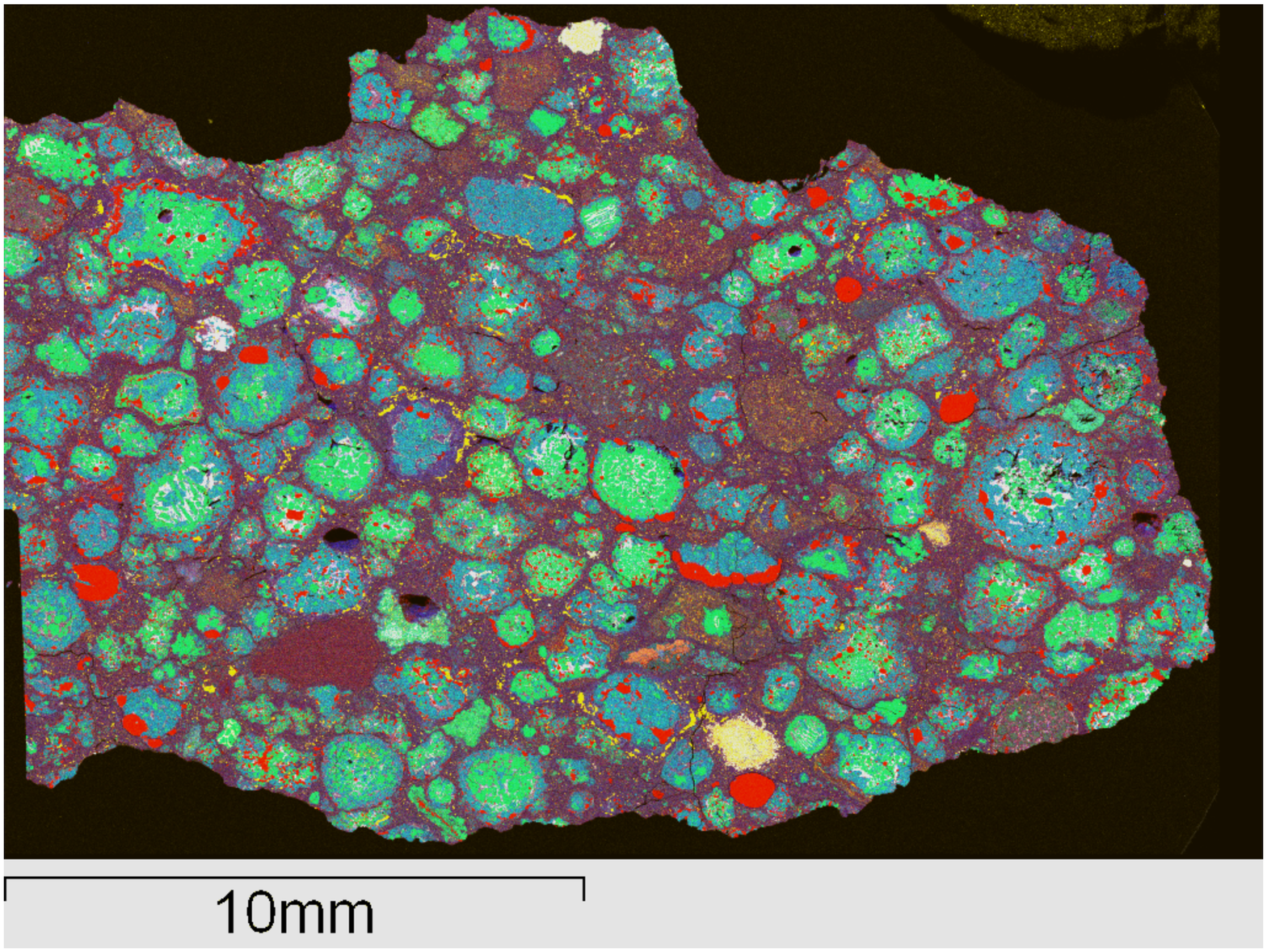}
   \caption{Composite false colour chemical map of the interior of the Renazzo chondrite: aluminum (white), magnesium (green), silicion (blue), calcium (yellow), iron (red). Chondrules are the abundant Mg- and Si-rich objects, CAIs are the rare yellow white inclusions and iron-nickel metal are the red grains. Credit: Anton Kearsley (NHM London).}
    \label{figsketch}
\end{centering}
\end{figure}

Some SLRs have been synthesised via the irradiation of the SPD gas/dust by proto-solar cosmic-rays \citep{Chaussidon2006}. As we saw in section 2 this is especially the case of $^{10}$Be ($T_{1/2}$ = 1.4 Myr) whose abundance in CAIs is variable \citep{Gounelle2013}. Independent evidence for early solar system irradiation is given by the variable and intense X-ray emission of protostars \citep{Feigelson2010}. Other SLRs such as $^{36}$Cl ($T_{1/2}$ = 0.3 Myr) or $^{41}$Ca ($T_{1/2}$ = 0.1	 Myr) could also have been produced by irradiation according to some models \citep{Gounelle2001,Duprat2007,Leya2003}. We note that the situation for these SLRs is complicated by the fact that their initial abundance in the solar system is not well constrained \citep{Liu2012}.

\subsection{Special role of $^{26}$Al and $^{60}$Fe}
Aluminium-26 ($T_{1/2}$ = 0.72 Myr) and iron-60 ($T_{1/2}$ = 2.6 Myr) are probably the SLRs, which have attracted the most attention during the last decade. Iron-60 is too rich in neutrons to be produced by irradiation processes \citep{Lee1998}. As $^{60}$Fe is absent from massive star winds, its only possible source in the early solar system is - at least one - supernova, which is known to produce large abundances of that SLR \citep{Wang2007}. Because the initial abundance of $^{60}$Fe has long been considered to be elevated ($^{60}$Fe/$^{56}$Fe $\sim$ 10$^{-6}$ \citep{Tachibana2003}), the presence of $^{60}$Fe has been interpreted as evidence for a nearby supernova \citep{Hester2004}. According to supernova nucleosynthetic models, the SN had to be located within a few pc from the nascent solar system \citep{Looney2006}. In N-body simulations of the birth environment of the Sun it is often assumed that the progenitor of this supernova was a star with a mass of $\approx$ 25 \Msun (see section 6). However, from the cosmochemical side there are some difficulties with this scenario.

\subsection{Cosmochemical constraints on the birth environment}
It is very unlikely to find a protoplanetary disc or a dense core that close to a SN. Before they explode as SNe, massive stars carve large ionized regions in the ISM (called HII regions) where the gas density is too low and temperature too high for star formation to take place. Observations show that even around a massive star that still needs to evolve for 2 Myr before it becomes a SN, discs and cores are found only several parsecs away \citep{Hartmann2005}, too far to incorporate  $^{60}$Fe at the solar abundance. In addition, because SNe ejecta are vastly enriched in $^{60}$Fe relative to $^{26}$Al and their respective solar abundances (Woosley \& Weaver 2007), all models relying on SN injection lead to a $^{26}$Al/$^{60}$Fe ratio far lower than the initial solar ratio, unless specific and ad hoc conditions are adopted \citep{Pan2012}. Finally, the initial abundance of $^{60}$Fe has been recently revised downwards by almost two orders of magnitude  \citep{Moynier2011,Tang2012}. 
The revised  $^{60}$Fe abundance  is
compatible with a {\bf somewhat} enhanced galactic background due to stars born and died in the same molecular cloud as the Sun but in a prior generation (Gounelle et al.,
2009).

\begin{figure}
\begin{centering}
\includegraphics[width=0.38\textwidth]{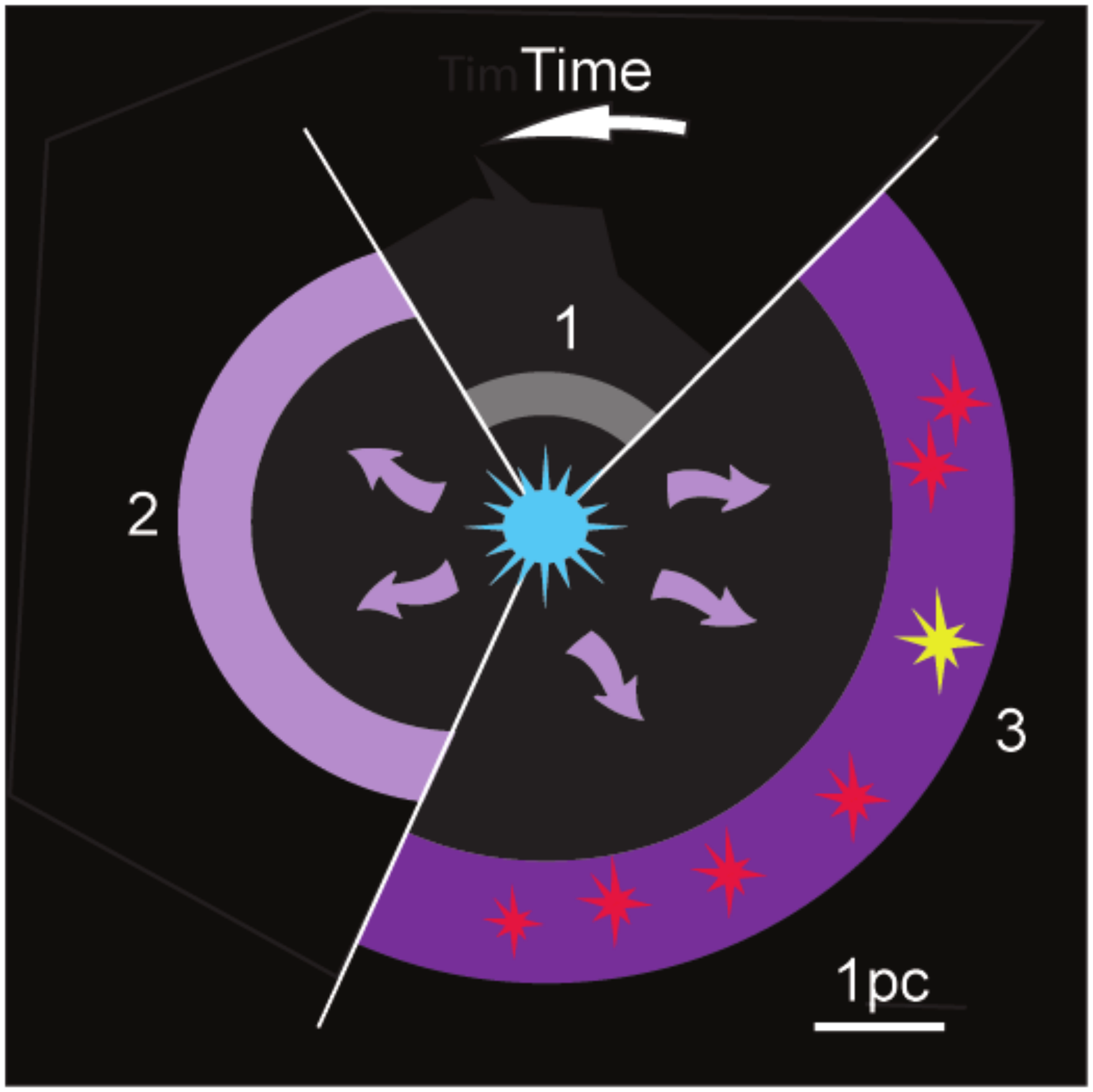}
   \caption{Incorporation of $^{26}$Al in a dense shell created by a massive star wind. Phases 1 and 2 correspond to the collection of interstellar gas and injection of $^{26}$Al by the wind (arrows). Phase 3 corresponds to the gravitational collapse of the shell and the formation of a new, $^{26}$Al-rich star and other stars including the Sun (yellow). The whole process lasts a few Myr (see text).}
    \label{x}
\end{centering}
\end{figure}

Because supernovae vastly overproduce $^{60}$Fe relative to $^{26}$Al, $^{26}$Al cannot originate from a single supernova as first proposed by \citet{Cameron1977}. It also implies $^{26}$Al cannot have a background origin because at the molecular cloud scale, SLRs are mainly due to supernovae. In addition, the $^{26}$Al distribution in the early solar system seems to have been heterogeneous \citep{Liu2012hib,2011ApJ...735L..37L}, which is incompatible with a global (molecular cloud) scale origin \citep{Young2014}.

The most likely origin for $^{26}$Al is therefore local, i.e. the wind of a single massive star \citep{Arnould2006}. \citet{Tatischeff2010} proposed that a single runaway Wolf-Rayet (WR) star injected $^{26}$Al in a shock-induced dense shell. Given the required velocity of the star (20 pc/Myr) and gas ambient density (n~$\sim$~100 cm$^{-3}$) it is however likely the star will have escaped the molecular cloud well before the star collects enough gas and enters the WR phase during which $^{26}$Al is injected. This model suffers from the brevity and therefore the rarity of the WR phase.  \citet{2012A&A...545A...4G} have proposed that $^{26}$Al originated in a wind-collected shell around a massive star (Figure 2). In this model, the dense shell collected by the massive star wind is continuously enriched by $^{26}$Al brought to the surface of the star by convection promoted by rotation \citep{Meynet2008}. After several Myr, the collected shell reaches densities large enough to make it gravitationally unstable and a new star generation forms in the shell. In this model, the Sun is a second generation star and the massive star having provided $^{26}$Al can be seen as its parent star. This mechanism is generic in the sense that this mode of star formation, though not the rule, is relatively common \citep{Hennebelle2009}. This model constraints the astrophysical context of our Sun's formation as it requires that the parent cluster to which belongs the parent star (baptised {\it Coatlicue}) contained roughly 1200 stars \citep{2012A&A...545A...4G}.

In conclusion, though the origin of SLRs is still a complex and debated issue (see Davis et al. 2014 for a more detailed review) it seems that three star formation spatial scales are relevant. SLRs with the longest half-lives originate at the scale of star complexes,
while $^{60}$Fe ($T_{1/2}$ = 2.6 Myr) originates at the scale of the molecular cloud and $^{26}$Al ($T_{1/2}$=
0.72 Myr) at the local massive star wind scale. The distribution of SLRs in the solar system
is the record of hierarchical (in termes of spatial and temporal scales) star formation in the ISM.

\section{Protoplanetary discs}

Another source of potential information about the formation of the solar system are young stars that are currently forming planets from the disc material surrounding them \citep{Safronov1969}. These
discs form as a consequence of angular momentum conservation during the star formation process \citep[e.g.][]{1987ARA&A..25...23S}. In the early stages of the star formation process the disc mediates accretion from the parental core to the forming protostar, while at later stages it provides the natural location and material for the planet formation process. The observational study of the properties and evolution of protoplanetary discs is thus a direct probe of the initial conditions for planet formation.

The presence of protoplanetary discs around young stars
was originally inferred from the emission in excess to the stellar photosphere at infrared- and submillimetre-wavelengths \citep[eg.][]{1986ApJ...309..755B,1990AJ.....99..924B}. It was very rapidly understood that the excess emission was likely due to a disc-like distribution of 
dust located around and heated by the young star at the centre, this interpretation was also consistent with the dynamics of the molecular gas, which could be interpreted as orbiting the star in Keplerian fashion. The inferred sizes (up to few times 100~AU in radius) and masses (up to 10-20\%\ of the central star mass) were also consistent with the expected 
values for a pre-solar nebula. These ideas were  then spectacularly confirmed by the HST silhouette optical absorption images a few years later \citep{1993ApJ...410..696O}. Following these initial observations, protoplanetary discs around young solar analogues have been extensively studied in nearby star forming regions. The recent review by \citet{2011ARA&A..49...67W} summarizes most of our current knowledge of the general properties of protoplanetary disc populations in nearby star forming regions.

\subsection{Protoplanetary disc lifetimes}

The current best estimate of the lifetime of the disc, or the timescale for planetesimal and gas giant formation, is derived 
from the variation of the fraction of stars, which show infrared excess as a function of the age of the star forming regions \citep[e.g.][]{2007ApJ...662.1067H}, which is also found to be consistent with the variation of the fraction of young stars that show signs of interactions 
with the inner disc \citep[e.g.][]{2010A&A...510A..72F}. The fraction of young stars with inner disc is found to drop exponentially with an e-folding time of $\sim$3~Myr. This estimate is uncertain because of a number of possible systematic and environmental effects, e.g. the uncertainties in the age determination of young stars \citep{2013MNRAS.434..806B}, or environmental disc dispersal mechanisms in clusters, which may not be applicable to the progenitors of the bulk of the field stellar population \citep{2014arXiv1409.0978P}. Nevertheless, it is obvious that if the majority of stars host planetary systems, the formation process has to occur within a few Myr, at least for planet types, which contain significant gaseous envelopes (gas giants, ice giants and low-density super-Earths). In large star forming complexes in the Magellanic Clouds and our own Galaxy, there is controversial evidence for stars with circumstellar discs at relatively old ages  \citep[up to and even beyond 30~Myr][]{2010ApJ...720.1108B,2011ApJ...739...27D}. While in nearby regions the few existing candidates have not been confirmed so far \citep{2013A&A...558A.114M}, it is possible that a small but non-negligible fraction of the stellar population born in large complexes reaccrete a disc and possibly have multiple chances of forming planetary systems \citep{2014A&A...566L...3S}.

\subsection{Protoplanetary disc gas content and chemistry}

Young protoplanetary discs are gas rich and most of the raw material available for planet formation is in the form of molecular gas, just like in clouds and cores. Most of this mass is in the observationally elusive form of H$_2$ and can only be traced, indirectly, through the emission of less abundant species. The molecular gas content, chemistry and emission from protoplanetary discs has been recently reviewed by \citet{2014arXiv1402.3503D}, and we refer to that work for a detailed account. The key feature for constraining the conditions for planet formation is that around young solar analogues most of the disc midplane is at low temperature and the most abundant molecular species (with the obvious exception of H$_2$) are frozen inside the ice mantles of grains. In these conditions, the chemistry is likely to be dominated by reactions on ices and may lead to significant production of complex organic and pre-biotic molecules, as observed in laboratory experiments \citep[e.g.][]{2002Natur.416..403M,2005ApJ...626..940H,2014A&A...561A..73I}. While complex organic compounds have been revealed in solar system objects \citep[][]{2006M&PS...41..889G,2009M&PS...44.1323E}, the direct detection and abundance measurements in protoplanetary discs is challenging. Nevertheless, in the cold disc midplane cosmic-ray induced desorption of ices may release a small fraction of the complex molecules in the gas phase, a process similar to the one that is thought to occur in cold protostellar cores
\citep{2012ApJ...759L..37C,2014ApJ...787L..33J}. 

Alternative promising locations to search for these complex compounds are the snowlines\footnote{The snow, or ice, line is the distance from a central protostar beyond which ice grains can form.} of the major molecular species, which could carry in the gas phase complex organic molecules as impurities as part of the sublimation process. The major snowlines that may produce such an effect are the CO snowline at $\sim$20K and the H$_2$O one at $\sim$150K. The CO snowline is now within grasp of the ALMA observatory in nearby star forming regions \citep{2013A&A...557A.132M,2013Sci...341..630Q}, and much work dedicated to the characterisation of this important transition in protoplanetary discs is expected in the coming years. Future instruments may help quantifying gas to dust ratio by combining high spatial and spectral resolution. 

\subsection{Protoplanetary disc masses and mass distribution}

As a consequence of the complexity of the molecular gas chemistry in discs and the difficulty in observing H$_2$, the most reliable mass estimates for protoplanetary discs still come from the measurements of the thermal dust emission. Most of the disc is optically thin at submm and longer wavelengths. Therefore, the thermal dust emission is directly proportional to the mass, which can be readily estimated from the 
observed fluxes assuming a dust opacity coefficient and a temperature structure for the disc. The latter is normally computed using a self consistent thermal balance model,  for typical accretion rates measured in young pre-main sequence objects the heating from the central star dominates the disc heating and the thermal structure of the dust is usually well constrained by models \citep[see e.g.][]{2007prpl.conf..555D,Bitsch+etal2014a}\footnote{The next challenge in modelling discs will be to include simulataneously an advanced radiation transport treatment and the three-dimenisonla dynamics.}
The disc masses derived with this method are typically between 0.2\%\ and 20\%\ of the mass of the central star, implying that a large fraction of young solar analogues are hosting a disc with a mass well above the minimum mass solar nebula and in principle capable of forming a planetary system similar to our own \citep[e.g.][]{2013ApJ...771..129A,2011ARA&A..49...67W}.

High angular resolution observations allow to resolve the disc emission and estimate the distribution of matter in the disc. The surface density of solids, traced by
the thermal dust emission, are typically found to fall off  as $r^{-1}$ with radial distance $r$ until a critical radius and then drop sharply, consistently with an
exponential fall off \citep[e.g.][]{2007A&A...469..213I,2008ApJ...678.1119H,2009ApJ...700.1502A,2013A&A...558A..64T}. The observed surface densities of the most massive discs are broadly consistent with the average surface density of our own solar system 
 (Andrews et al. 2009). However, most systems that have a total mass above the minimum mass solar nebula threshold are larger than our own solar system, resulting in a lower
surface density.
These estimates are not without uncertainties: firstly, the solids are assumed to trace approximately 1\%\ of the total mass (following the estimates of the dust to gas ratio in the interstellar medium), however this number is highly uncertain in discs and also expected to evolve with time and location, as gas and dust evolve following different pathways; secondly, the dust opacity coefficient depend on the composition, size, shape, and porosity of the dust grains \citep{2004ASPC..323..279N,2014arXiv1402.1354T}. Measurements of proto planetary disc masses are relatively rare as of today: up-coming ALMA observations may be able to provide a statistical view of the distribution of dust surface densities and possibly also gas surface densities.

\subsection{Dust evolution and the first steps of solids growth}

As the dust emission is optically thin, multi-wavelength observations at submillimetre wavelength allow us to probe the 
dust opacity coefficient, which in turn can be a powerful probe of the grain population properties. In a protoplanetary
disc, grains are expected to grow and settle onto the midplane \citep[for a recent review, see][]{2014arXiv1402.1354T}. As grains grow 
to sizes of the same order or larger than observing wavelengths,  the dust opacity coefficient dependence 
with wavelength changes from the typical value of the sub-micron size interstellar grains ($\kappa_\nu\sim\nu^{-1.7}$) to a much shallower
dependence, approaching $\kappa_\nu\sim const.$ in the limit of a population of grains all much larger than the observing wavelength 
\citep{1986ApJ...309..755B,2006ApJ...636.1114D,2007prpl.conf..767N}. This argument has been used to constrain the level of grain growth 
in protoplanetary discs, finding that in most discs  around young stars the dust has grown to pebble-size aggregates \citep{2001ApJ...554.1087T,2003A&A...403..323T,2005ApJ...626L.109W,2010A&A...512A..15R}.

The observational results can be understood in the framework of global dust evolution models in discs \citep[e.g.][]{2010A&A...516L..14B}, which 
describe the evolution of dust in the disc environment based on constraints from laboratory measurements of the grain-grain 
collision outcomes \citep[see e.g.][]{2008ARA&A..46...21B}. More specifically, the emerging observational constraints on the dust grain 
size distribution as a function of disc radius \citep{2011A&A...529A.105G,2012ApJ...760L..17P,2013A&A...558A..64T} matches the 
expectations from the evolutionary models \citep{2012A&A...539A.148B,2014MNRAS.437.3037L}. 

This apparent success hides a number of difficulties that still need
to be solved, in particular, smooth disc models still predict a radial drift and fragmentation of the large grains that is inconsistent with the
observations and require to artificially slow down the radial drift. The planet formation process then requires that additional growth barriers are overcome, a variety of possibility exist, from pressure traps generated by disc instabilities \citep[e.g.][]{2012A&A...538A.114P}, localised growth across snowlines \citep{2013A&A...552A.137R}, or the efficient accretion of pebbles onto existing planetary cores
\citep{2014arXiv1408.6094L}. Some of these models predict observational signatures that are within reach 
of the current generation of observing facilities. Indeed the effect of pressure traps induced by planet-disc interaction has been observed in some systems with ALMA \citep[e.g.][]{2013Sci...340.1199V}. 

An intriguing development in recent years has been the observation by several groups of 
the possibility of significant dust evolution before the disc stage in prestellar cores and protostars \citep{2012ApJ...756..168C,2014A&A...567A..32M}. If confirmed, these findings would imply a significant revision of our 
assumptions on the initial conditions for planet formation in protoplanetary discs.

\section{Planet growth}

\subsection{From dust to pebbles}

Planets from the gas and dust present in protoplanetary discs.  In the first stage of planet formation dust and ice
particles collide and stick together to form larger aggregates
\citep{DominikTielens1997}. The aggregates maintain a high porosity during the
growth, since collision speeds are initially too low to cause restructuring.
 Particles obtain increasingly higher collision speeds as they grow in size and
their frictional coupling with the gas diminishes (Voelk et al., 1980; Ormel \&
Cuzzi, 2007), so that particle pairs decouple from the smallest eddies and thus
will have different velocity vectors even when at the same location. An additional contribution to the
collision speed is the relative radial and azimuthal drift between
different-sized particles \citep{Weidenschilling1977a}.

Silicate dust aggregates in the inner protoplanetary disc are compactified by
collisions when they reach approximately mm sizes \citep{Zsom+etal2010}. The
combination of high collision speeds and low porosity halts growth by direct
sticking, since compact particles cannot dissipate enough energy during a
collision to allow sticking \citep{Guettler+etal2010}. This is referred to as
the {\it bouncing barrier}. Ice monomers are generally stickier than silicate
monomers and this increases their resistance against compactification
\citep{Wada+etal2009}. The size of the constituent monomers also plays an
important role in determining the collision outcome. Ice aggregates, which
consist of very small monomers of $0.1$ $\mu$m sizes, can stick at up to 50 m/s
collision speeds \citep{Wada+etal2009}. It is nevertheless not known whether
such small ice monomers are prevalent in protoplanetary discs or whether
sublimation and condensation cycles drive ice particles to much larger sizes
\citep{2013A&A...552A.137R}.

\subsection{From pebbles to planetesimals}

The continued growth from mm-sizes faces the formidable {\it radial drift
barrier}. The gas in the protoplanetary disc is slightly pressure supported in
the radial direction, since viscous heating and stellar irradiation heat the
inner regions more strongly \citep{Bitsch+etal2014a}. This outwards-directed
push on the gas causes the gas to orbit slower than the Keplerian speed, by
approximately 50 m/s \citep{Weidenschilling1977a}. Solid particles do not sense
the pressure difference from front to back (since their material density is
much higher than the gas density), so particles would orbit at the Keplerian
speed in absence of gas drag. However, the drag from the slower moving gas
drains particles of their angular momentum and causes them to spiral inwards
towards the star. In the asteroid belt the radial drift peaks for m-sized
particles, which fall towards the star in a few hundred years, to be destroyed
at the silicate sublimation line close to the star. The peak of the radial
drift occurs for cm-sized particles in the outer protoplanetary disc where the
giant planets form.

There are three main ways by which particles can cross the radial drift barrier:

{\it Mass transfer}. At collision speeds from 1 to 25 m/s particles can grow by
mass transfer. The projectile is destroyed in the collision but leaves up to
50\% of its mass attached to the target \citep{Wurm+etal2005}. Particles stuck
at the bouncing barrier do not collide at such high speeds. However,
artificially injected cm-sized seeds can grow from the sea of bouncing barrier
particles through mass transfer \citep{Windmark+etal2012a}. The growth rate is
nevertheless too low to compete with the radial drift, so formation of
planetesimals by direct sticking requires a reduction in the radial drift speed
of the particles, e.g.\ through the presence of a long-lived pressure bump.
Such a pressure bump may arise at the inner edge of the dead zone
\citep{Lyra+etal2008b,Kretke+etal2009,Drazkowska+etal2013} or around the water
ice line where the jump in dust density has been proposed to cause a jump in
the turbulent viscosity \citep{KretkeLin2007}, although this requires a very
steep viscosity transition at the ice line \citep{Bitsch+etal2014b}.

{\it Fluffy particles}. Ice aggregates consisting of small monomers acquire
very low densities during their growth, down to $10^{-5}\,{\rm g\,cm^{-3}}$
\citep{Okuzumi+etal2012}. Particle growth is only able to outcompete the radial
drift if the particle grows larger than the mean free path of the gas molecules
and enters the Stokes drag force regime. Here the friction time is proportional
to the particle radius squared; hence decoupling is much more rapid than in the
Epstein regime valid for small particles \citep{Johansen+etal2014}. 
Very fluffy particles are large enough to be in the Stokes regime and can
cross the radial drift barrier within 10 AU by growing in size while drifting
radially \citep{Okuzumi+etal2012}.
 Hence this is a way to
cross the radial drift barrier, if particles can remain fluffy and avoid
compactification and erosion by the gaseous headwind.

{\it Particle concentration}. Particles can become strongly concentrated in the
turbulent gas, triggering the formation of planetesimals by a gravitational
collapse of the overdense regions. Particles concentrate either {\it passively}
or {\it actively}. For passive concentration the general mechanism is that
particles pile up where the gas pressure is high, since any rotating structure
in the turbulent gas flow must be in semi-equilibrium between the dominant
forces, namely the pressure gradient force and the centrifugal/Coriolis force (Barge 
\& Sommeria 1995).
Particles do not sense gas pressure and must therefore move towards the
direction of increasing gas pressure. On the smallest scales of the turbulent
flow, approximately km under prevalent conditions in the asteroid belt,
mm-sized particles pile up between rapidly over-turning eddies
\citep{Cuzzi+etal2001,Cuzzi+etal2008}. The largest scales of the protoplanetary
disc are dominated by the Coriolis force, and the turbulence organises into
axisymmetric pressure bumps, surrounded by zonal flows
\citep{FromangNelson2005,Johansen+etal2009a,Simon+etal2012}, or elongated
vortices \citep{KlahrBodenheimer2003,LesurPapaloizou2010,LyraKlahr2011}. In the
streaming instability scenario the particles actively drive concentration, by
piling up in filaments, which locally accelerate the gas towards the Keplerian
speed and hence do not drift radially
\citep{YoudinGoodman2005,JohansenYoudin2007,BaiStone2010a}.  Strong particle
concentration by the streaming instability is triggered at a metallicity
slightly higher than solar \citep{Johansen+etal2009b,BaiStone2010b}. The
gravitational collapse of overdense regions leads to the formation of
planetesimals of sizes from 100 to 1000 km 
\citep{Johansen+etal2007,Johansen+etal2011,Johansen+etal2012,Kato+etal2012}.

The solar system contains remnant planetesimals from the epoch of planet
formation, in the asteroid belt and in the Kuiper belt. These populations of
planetesimals can be studied to infer constraints on the actual processes that
led to planetesimal formation in the solar system. The relative lack of
asteroids smaller than 100 km in diameter, compared to a direct extrapolation
from the largest sizes, does not agree with the formation of asteroids by
hierarchical coagulation in a population of km-sized planetesimals \citep[][but
see \citet{Weidenschilling2011} for an alternative view]{Morbidelli+etal2009}.
Asteroids may instead have had birth sizes larger than 100 km, in agreement
with gravitational collapse models, while the smaller asteroids that are found
in the asteroid belt today are mainly collisional fragments
\citep{Bottke+etal2005}. Large birth sizes have also proposed to explain a
similar lack of small Kuiper belt objects \citep{SheppardTrujillo2010}.

\subsection{From planetesimals to planets}

Planetesimals are the building blocks of both terrestrial planets and the cores
of the giant planets. In the {\it core accretion} scenario giant planets form
as gas from the protoplanetary disc collapses onto a core that has grown to
approximately 10 Earth masses by accumulation of planetesimals
\citep{Mizuno1980,Pollack+etal1996}. In this picture, the ice giants Uranus and
Neptune only manage to attract a few Earth masses of gas before the gaseous
protoplanetary disc dissipates after a few million years. Core accretion by
planetesimals is nevertheless very slow because the number density of
planetesimals is low in the region where the giant planets form.
An enhancement of 4-6 over the Minimum Mass Solar Nebula is needed to form
Jupiter and Saturn within 10 million years in the classical core accretion
scenario \citep{Pollack+etal1996}.
\begin{figure}
\begin{centering}
 \includegraphics[width=8cm]{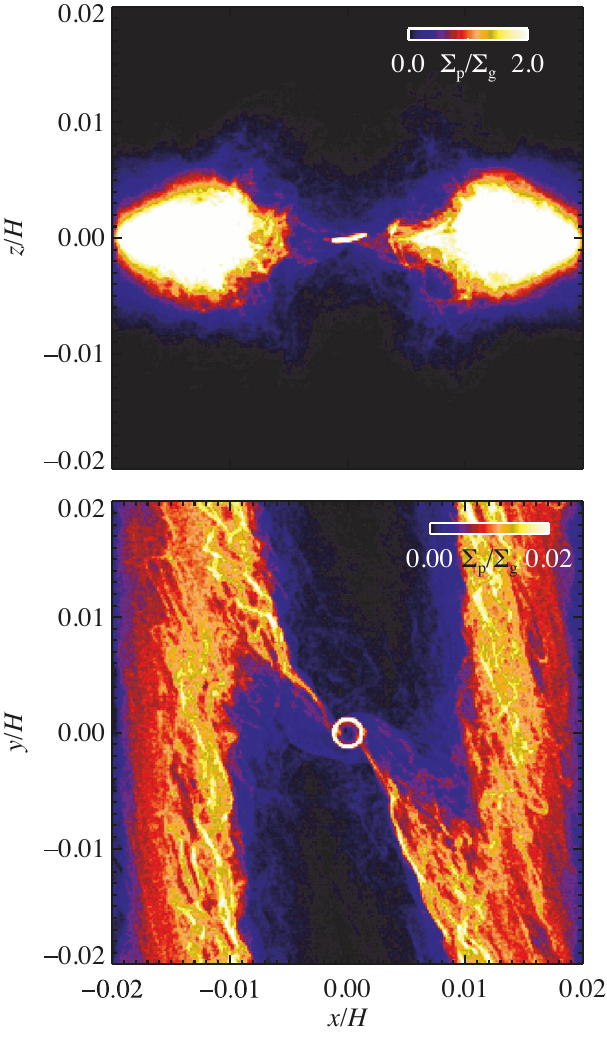}
  \caption{Accretion of pebbles onto a 1000-km-scale protoplanet. The
  simulation box corotates with the protoplanetary disc at an arbitrary
  distance from the star. The top plot shows the column density of cm-sized
  pebbles in the radial-vertical plane, while the bottom plot shows the column
  density in the radial-azimuthal plane. Two streamers of material enter the
  Hill sphere, which is approximately 0.007 times the gas scale-height $H$ for
  the considered protoplanet mass, and feed a particle accretion disc orbiting
  the protoplanet. Figure adapted from \cite{JohansenLacerda2010}.}
  \label{f:Figure6}
\end{centering}
\end{figure}

Core accretion timescales can be decreased when accreting planetesimal {\it
fragments}. Small fragments have their scale heights damped by gas drag and
hence the growing core can accrete a much larger fraction of the solid material
\citep{Goldreich+etal2004,Rafikov2004}. However, global disc simulations show
that a large fraction of the slowly drifting fragments are trapped in
resonances with the cores \citep{Levison+etal2010}. The formation of a system
of giant planets is further complicated by the excitation of planetesimal
eccentricities and inclinations when scattered by the growing cores. This leads
to a slow oligarchic growth phase with growth rates that are too low to form
cores within the life-time of the protoplanetary disc \citep{Levison+etal2010}.

Pebbles left over from the planetesimal formation process can be accreted very
efficiently by the growing cores
\citep{JohansenLacerda2010,OrmelKlahr2010,LambrechtsJohansen2012,MorbidelliNesvorny2012}.
The Hill sphere of the core denotes the radial distance over which an incoming
particle on a faster (interior) or slower (exterior) orbit is scattered
gravitationally by the core. The core only extends a small fraction of its Hill
radius (0.1\% at 5 AU, 0.01\% at 50 AU). Gravitational focussing of the
incoming planetesimals makes the accretion radius much larger than the physical
radius, but planetesimals are still only accreted from about 3\% of the Hill
radius at 5 AU. Pebbles of mm-cm sizes experience strong drag during the
gravitational scattering and lose enough energy to be gravitationally bound to
the core, for any impact parameter up to approximately the Hill radius (see
Figure \ref{f:Figure6}). This leads to very high accretion rates, about 1,000
times faster than classical core accretion at 5 AU and 10,000 times faster at
50 AU. Such high growth rates are needed to explain the formation of gas giants
observed in wide orbits around some young stars \citep[e.g.\
HR8799,][]{Marois+etal2008,Marois+etal2010} assuming they are not formed from other means (like gravitational instability). As detailed in section  4.4, protoplanetary discs observed at
mm- and cm-wavelengths show signs of large populations of mm-cm-sized pebbles,
which can boost planet formation \citep{2003A&A...403..323T,2014arXiv1402.1354T}.


Many of the young stars surrounded by discs are part of a cluster of  stars, indeed it seems that most stars are born in a clustered environment (Lada \& Lada 2003).
In the next section we will discuss the possible birth environment of the Sun. 
In particular we consider whether we can constrain the properties
of the birth cluster, indeed its possible identity, and whether
we  can  identify solar siblings, i.e.\ the
stars that were born from the same molecular cloud as the Sun
\citep[see for example,][]{2009ApJ...696L..13P, 2010ARA&A..48...47A,2013A&A...549A..82P}.

\section{Formation within stellar clusters}

\subsection{Evidence suggesting the Sun formed in a stellar cluster}

Analysis of the isotopic abundances of meteorites reveal that they
contain the decay products of the radioactive isotopes $^{26}$Al and
$^{60}$Fe, which have half lives of 0.7 Myr and 2.6 Myr
 \citep{1976GeoRL...3..109L}.  As mentioned in section 3 the most likely source of these
isotopes are supernovae explosions \citep[e.g.][]{2000ApJ...538L.151C}. How many is still a matter of debate. Given the observed distribution of stellar masses of newly-formed stars follows
a power law favouring less-mass stars
 \citep[$dN/dm \propto m^{-2.35}$ - ][]{1955ApJ...121..161S}
 one would need a fairly large group of stars reasonably close to the forming Sun to
explain the observed abundancies. 

There are other indications that the Sun once was part of a
fairly dense and therefore large stellar cluster. First, the fairly abrupt cut-off in the mass distribution
at 30-50AU, which is usually not observed in protoplanetary discs, but could be caused
by a fly-by during the protoplanetary disc phase. Second, the trans-Neptunian objects Sedna and 2012 VP$_{113}$ (Morbidelli \& Levison 2004, Kenyon \& Bromley 2004, Brasser et al. 2006, Trujillo \& Sheppard 2014) with their high excentricities are another property that could be explained by an encounter during the planet forming phase. Calculations of encounter probabilities in clusters suggest that the Sun was formed in a stellar cluster
containing 2000 to 10000 stars (Portegies Zwart 2009). Further evidence
for the sun having been in a cluster, comes from the measured tilt of the sun's rotation, one suggestion
being that the protoplanetary disc was titled by a star-disc interaction (Heller 1993;  Thies et al. 2005).

\subsection{Encounters and other processes within stellar clusters}

Stellar clusters are potentially dangerous environments for planetary
systems. Close encounters between stars are relatively frequent in
such crowded places. The timescale for a particular star to have
another star pass within some distance $R_{\rm min}$ can be
approximated by 
\citep{2008gady.book.....B}

\begin{equation}
\tau_{enc}  \simeq  3.3 \times 10^{7} {\rm yr} \left( {100 \ {\rm pc}^{-3} \over
n } \right) \left( { v_\infty \over 1 \ {\rm km/s} } \right) \nonumber 
\left( { 10^3 \, {\rm AU} \over r_{min} } \right) \left( { {\rm
M}_\odot \over M} \right) 
\label{mbd_spz_equation1}\end{equation}

One can distinguish between different phases during which such an encounter could happen: (i) the early phase when the star is surrounded by an accretion disc (first few Myr), (ii) the planet growth phase, and (up to 100 Myr), and (iii) after the solar system was fully formed (up to now).
Close encounters occurring whilst stars still possess protoplanetary
discs could lead to their truncation
 \citep{2001Icar..153..416K,2009MNRAS.400.2022F,2014A&A...565A.130B,2014MNRAS.441.2094R}.
 In addition, illumination by massive stars may lead to
the early evaporation of protoplanetary discs around young stars
\citep{2000A&A...362..968A}.
Thus, clusters may inhibit the planet formation
process.

Planetary systems, once formed, are also affected by other stars
within clusters.  Fly-by encounters with other stars may perturb
otherwise stable planetary systems 
 \citep[e.g.][]{2007MNRAS.378.1207M,2011MNRAS.411..859M}
Changes in planetary orbits within a system can lead to the growth
in eccentricities until orbits cross.  Scattering can then lead to the
ejection of planets leaving those remaining on more bound and
eccentric orbits. For lower-mass planets on tighter orbits, orbit
crossing tends to lead to collisions between planets
\citep[e.g.][]{2013arXiv1311.6816D}.
Alternatively, stars hosting planetary systems may exchange
into (wide) binaries. The stellar companion may then perturb planets'
orbits via the Kozai-Lidov mechanism, where planets' orbits
periodically pass through phases of high eccentricity. Planetary
orbits may then cross leading to scattering or collisions
\citep[e.g.][]{2007MNRAS.377L...1M,2013arXiv1311.6816D}. In the early phases the probability of encounters are highest, as afterwards clusters expand  and therefore the encounter likelihood decreases. 

\subsection{Placing the solar system inside a stellar cluster}

We have seen above how the measured isotopic abundances in meteorites
are well explained by the enrichment of the protosolar-system material
from a nearby supernova formed from a relatively massive star or from a combination of supernovae and a massive $^{26}Al$-producing star. The
rarity of such objects in turn led us to argue that the solar system
likely formed within a stellar cluster containing at least 2000 stars.
As just discussed, stellar clusters are hazardous environments.  The
rate of destructive close stellar encounters tends to increase with
increasing cluster mass (see Eq.\,\ref{mbd_spz_equation1}). Therefore,
there is an optimum stellar cluster containing around a few thousand
stars where pollution from supernovae is at least possible whilst at
the same time a reasonable fraction of planetary systems may survive
unperturbed
\citep{2001Icar..150..151A}.
 One can perform N-body
simulations of stellar clusters containing one 25 M$_\odot$ star to
determine the number of solar-like stars, which are 0.1 - 0.3 pc from
the 25 M$_\odot$ star when it explodes as a supernova. One can then
follow the subsequent trajectories of these polluted stars within the
cluster to determine the fraction of them, which avoid both perturbing
fly-by encounters and exchange encounters into binaries. Such
numerical studies of clusters containing 2100 stars (including one 25
M$_\odot$ star) reveal that some 25 percent of clusters contain
enriched, unperturbed solar-like stars (i.e.\ G-dwarfs), and usually
only one or two per cluster out of a total of 96 G-dwarfs 
\citep{2014MNRAS.437..946P}.
Therefore, roughly one percent of G-dwarfs from such
clusters are enriched whilst being unperturbed.

\subsection{M 67 as a possible host cluster}

If the Sun was indeed born in an open cluster, it is rather natural to
ask whether this cluster is still around today, and if so, whether we can identify it.
There have been a number of observations of star clusters of different ages dedicated to find planets in them (Meibom et al. 2013, and references therein). However, planet detection in clusters provides some added difficulties, for example, even nearby clusters are often more distant than the field stars one finds usually planets around), thus the frequency of planets in cluster environments compared to that in the field is still an open question. However, probably most clusters in the solar neighbourhood  loose a considerable fraction ($>$ 70\%) of their stars (Pfalzner \& Kaczmarek 2013).
Measuring detailed abundance patterns in stellar atmospheres through
high-resolution spectroscopy offers us the hope of identifying other
stars, which originate from the same gas cloud as the Sun
\citep[e.g.][]{2009ApJ...696L..13P, 2010ARA&A..48..431P}.
Even if
the cluster has long since dissolved, stars from it will still share
similar orbits whose properties could be measured astrometrically
\citep[e.g.][]{2010MNRAS.407..458B}.

The open cluster M 67 is one candidate which we will consider here, though there must have been
a number of cluster produced in the Milky way having similar ages and galactocentric radii. M 67 has an age. of about 3.5--4.8 Gyr
 \citep{2008A&A...484..609Y}
and the stars within it also have a composition similar to
the Sun. 
\citet{2011A&A...528A..85O}
 have identified one star in the
cluster, which is a better match to our Sun than most solar-like stars
in the solar neighbourhood. The analysis of 13 additional stars in M
67 confirm that the abundances of the Sun and M 67 are 
similar
\citep{2014A&A...562A.102O}.

\begin{figure}
\centering
\includegraphics[angle=90,width=12cm]{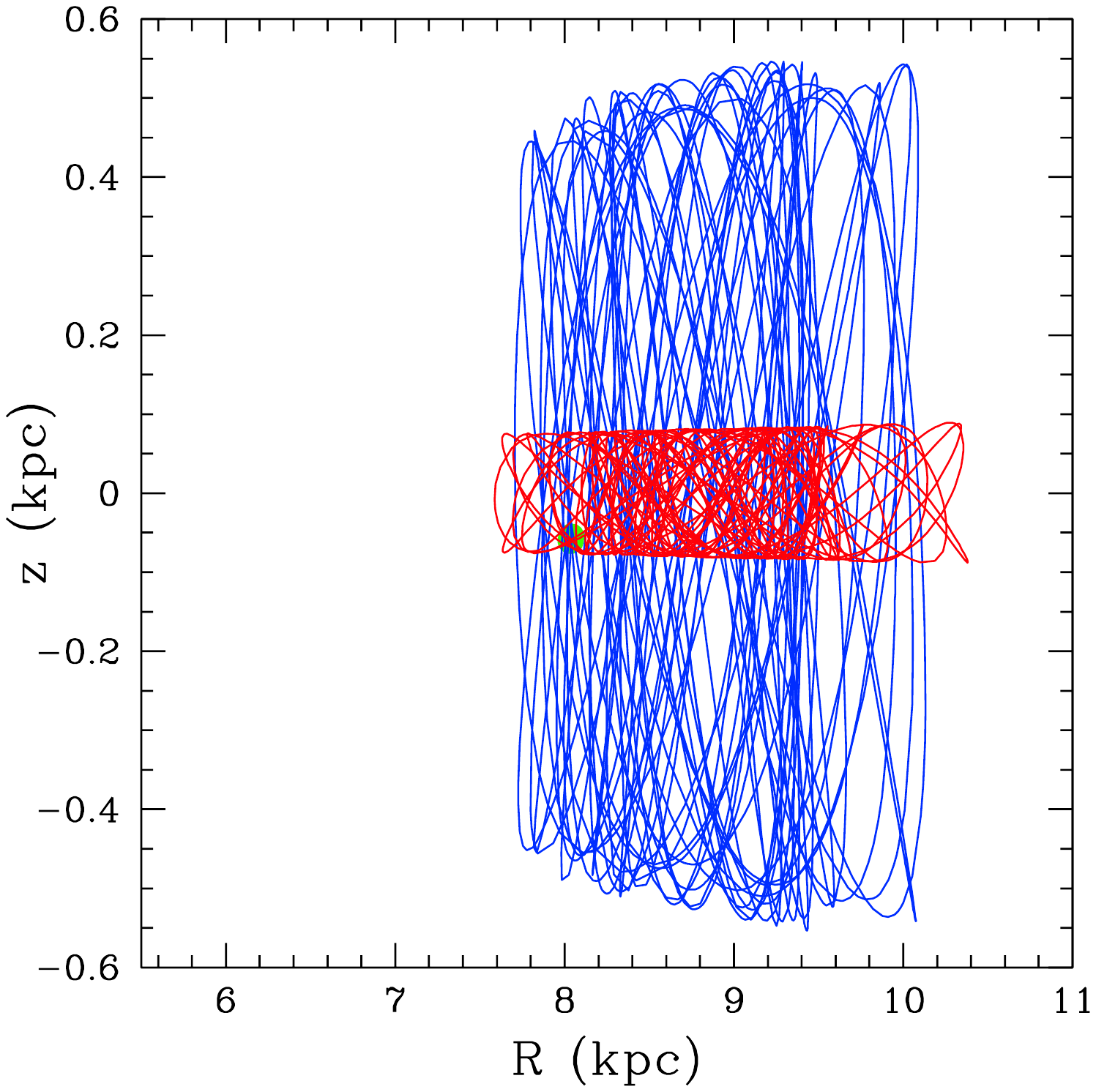}
\caption{The current orbits of the Sun (red line) and the open cluster
  M 67 (blue line) in cylindrical coordinates where $R$ is the
  cylindrical radius from the Galactic centre and $z$ is the distance
  from the Galactic plane 
  \citep[Figure 7 of][reproduced with permission of the AAS]{2012AJ....143...73P}.}
\label{mbd_spz_figure1}
\end{figure}

However, as pointed out by 
\citet{2012AJ....143...73P}
the current orbits of M 67 and the Sun are very different, as shown in
Fig.\ \ref{mbd_spz_figure1}, where we plot their current orbits in
cylindrical coordinates 
\citep[figure 7 from][]{2012AJ....143...73P}.
The
radial distance from the galactic centre, $R$, for the Sun and M67 are
similar. Earlier mention of abundance enhancements in the Sun
compared to its distance from the Galactic centre 
\citep{2009A&A...501..941H}
would indicate that the Sun experienced a radial migration of
0.8--4.1\,kpc 
\citep{2013A&A...558A...9M}. 
In that case M 67 should also have
migrated over the same distance. Possibly, the migration happened
while the Sun was still a member of its birth cluster. Extensive numerical analysis of the
orbit of the Sun in the Galactic potential however indicates that
this radial migration in the current Galactic potential would be
negligible 
\citep{Martinez-Barbosa_MNRAS}.

One can see from Fig.\ \ref{mbd_spz_figure1}  that
the orbit of M 67 takes the cluster significantly 
further out of the galactic plane to much larger values of $z$ than is experienced by the Sun. 
If one assumes that M 67 has always had its current orbit and that the Sun was ejected from M 67, 
this would imply that the Sun received a significant kick in order to leave it on its current orbit,
residing in the plane of the galaxy
 \citep{2012AJ....143...73P}.
 The point here is that it is not possible
to impart such a large kick whilst keeping the solar system intact.

A possible solution to this problem has been suggested in 
\citet{2014gustafssonetal}.
The key idea is that the orbit of M 67 has changed over
time. The orbit could originally have been more in the galactic plane
(i.e.\ with lower maximum values of $z$) allowing the Sun to escape
from it with a relatively low velocity. Scattering off giant molecular
clouds (GMCs) could then leave the cluster on the orbit we see today.
Simulations of scattering due to GMCs show that it is possible to put
M 67 on its current orbit beginning with an orbit similar to the Sun's
\citep{2014gustafssonetal}.
 In that case the Sun must have left M 67
before the scattering event with a GMC took place.

\section{The Kuiper Belt and the Late Heavy Bombardment}

\subsection{The Kuiper Belt}

Beyond Neptune lies a reservoir of remnant icy planetesimals from the
formation of the solar system analogous to the main asteroid belt
between Mars and Jupiter.  Known as the Kuiper belt, this distant
population displays an intricate orbital structure that has inspired
and constrained a number of ideas on the early evolution of our
planetary system. The high abundance of Kuiper belt objects (KBOs) in
mean-motion resonances with planet Neptune led to the idea that the
ice giant has migrated through the original disc of planetesimals,
sweeping a significant fraction into stable resonant orbits
\citep{1995AJ....110..420Malhotra}. The superposition of dynamically
hot (high inclination and eccentricity) and cold (low inclination and
nearly circular) orbits suggests that while some KBOs have been
scattered onto their current orbits \citep{2003Icar..161..404Gomes},
likely by the outer planets, others have formed in-situ and remained
unperturbed \citep{2011ApJ...738...13Batygin}. 
An alternative explanantion is provided by an encountering star that
scattered the Kuiper belt (Ida et al. 2000).  The current
morphology of the scattered Kuiper belt could only be reproduced if the mass
of the encountering star $M$ had an impact parameter of $b = 170 +
45(M/M_\odot - 0.5)$\,AU (Punzo et al. 2014).  However, such an
encounter would like scatter the entire Kuiperbelt. In addition, cooling off
part of the hot population to re-populate the cold population would require
some 1000 Pluto-mass objects, which are not observed.

\subsection{Early Dynamical Evolution}

The need to explain the dynamical structure of the Kuiper belt and to
reconcile the orbits of the giant planets with the discovery of
hot-Jupiters were main drivers in the evolution of planet migration
theories. In the solar system, the current paradigm considers that the
giant planets emerged from the protoplanetary disc in a more compact,
near-resonant configuration, all within 15 AU from the Sun
\citep{2010ApJ...716.1323Batygin}.  Angular momentum exchanges with
the surrounding disc of planetesimals led to the outward migration of
Neptune, Uranus, and Saturn, as they scattered small bodies inwards,
while Jupiter, sufficiently massive to eject planetesimals from the
solar system, migrated predominantly towards the Sun
\citep{1984Icar...58..109Fernandez, 1999AJ....117.3041Hahn}. The
diverging migrations of Jupiter and Saturn eventually resulted in a
mutual resonance crossing and triggered a dynamical instability
throughout the solar system, which impulsively expanded the orbits of
Uranus and Neptune to near their current locations and violently
scattered the planetesimal disc \citep{2005Natur.435..459Tsiganis}.
The Late Heavy Bombardment was also likely caused by this instability
event \citep{2005Natur.435..466Gomes}.

\subsection{The Late Heavy Bombardment}

Geochronological studies of the lunar rocks returned by the Apollo
missions yielded the surprising result that all large multi-ring
basins on the Earth's moon -- they are also called Mare: Orientale,
Crisium, Imbrium, Nectaris, Humorum, Serenitatis --  originated by
giant impacts in an extremely confined time interval between 4.0 and
3.7 Gyr ago \citep{1974E&PSL..22....1Tera, 1977PhChE..10..145Turner,
1974LPSC....5.1419Jessberger}. This process is called the Lunar (or
Late) Heavy Bombardment (LHB), as it was completed only 800 Myr after
solar system formation. It requires that long after final formation of
the terrestrial planets, there was an increased flux of small bodies
(up to about 100 km in size) in the inner solar system 3.8 Gyr ago.
Possible evidence of the LHB is also inferred from HED meteorites
\citep{1995Metic..30..244Bogard, 1995P&SS...43..527Kunz,
2003M&PS...38..669Bogard}, which are inferred to come from Vesta, and
Martian meteorites like ALHA84001 \citep{1996Natur.380...57Ash,
1997GeCoA..61.3835Turner}, although this is not undisputed
\citep{1999M&PS...34..451Bogard}.

Two general scenarios appear plausible for the early solar system
impact history: 

i) The LHB was the final phase of terrestrial planet accretion -- this
requires that the small body flux must have been considerably higher
before (implying somewhat unrealistic total mass estimations of
initial small body populations), and that the record of earlier
impacts must have been erased by the late impact phase 3.8 Gyr ago.

ii) The LHB was an episodic spike of the influx of asteroidal or
cometary bodies, caused by dynamical excitation of small body
populations in the asteroid or Kuiper belts.  As general explanation
for the dynamical excitation, planetary migration processes are
promising candidates, in particular the aforementioned resonance
crossing of Jupiter and Saturn \citep{2005Natur.435..466Gomes,
2005Natur.435..459Tsiganis}, i.e. in more general terms, the giant planet instability (Morbidelli et al., 2007, Levison et al., 2011).  Advanced models consider a more
prolonged bombardment history induced by planet migration processes
(Morbidelli et al. 2012).

However, there is increasing evidence that before the time of the LHB
3.8 Gyr ago, there were other episodic times of increased impact rates
on the Earth-moon system, registered by, e.g.\ Ar-Ar and zircon
chronology \citep{2013M&PS...48..241Fernandes,2008GeCoA..72..668N}, and also by other
meteorite parent bodies, e.g.\ 4.2 Gyr impact events affecting the LL
chondrite parent body \citep{1989Metic..24R.332Trieloff,
1994Metic..29Q.541Trieloff, 2004GeCoA..68.3779Dixon}. It was suggested
\citep{2014Trieloff} to consider close stellar encounters as a
possible reason for repeated episodic dynamical excitation.
However, there is increasing evidence that before the time of the LHB 3.8 Gyr ago, there were other episodic times of increased impact rates on the Earth-moon system, with concomitant impact episodes on other inner solar system bodies. For example, both Ar-Ar and zircon chronology (Fernandes et al. 2013, Nemchin et al. 2008) evidence an increased impact cratering rate 4.2 Ga ago, which was also experienced by the LL chondrite parent body (Trieloff et al. 1989, 1994, Dixon et al. 2004). Although the chronological data base has still to be improved, age clustering can only hardly be reconciled with stochastical bias effects, as multiple celestial bodies were affected.  As giant planet instabilities can only once trigger bombardments, it was suggested (Trieloff 2014) to consider close stellar encounters as a possible reason for repeated episodic dynamical excitation. However, the corresponding dynamical scenarios are only poorly explored. 

\subsection{Decoding KBO Diversity}

\begin{figure}
\begin{centering}
\includegraphics[angle=0,width=8cm]{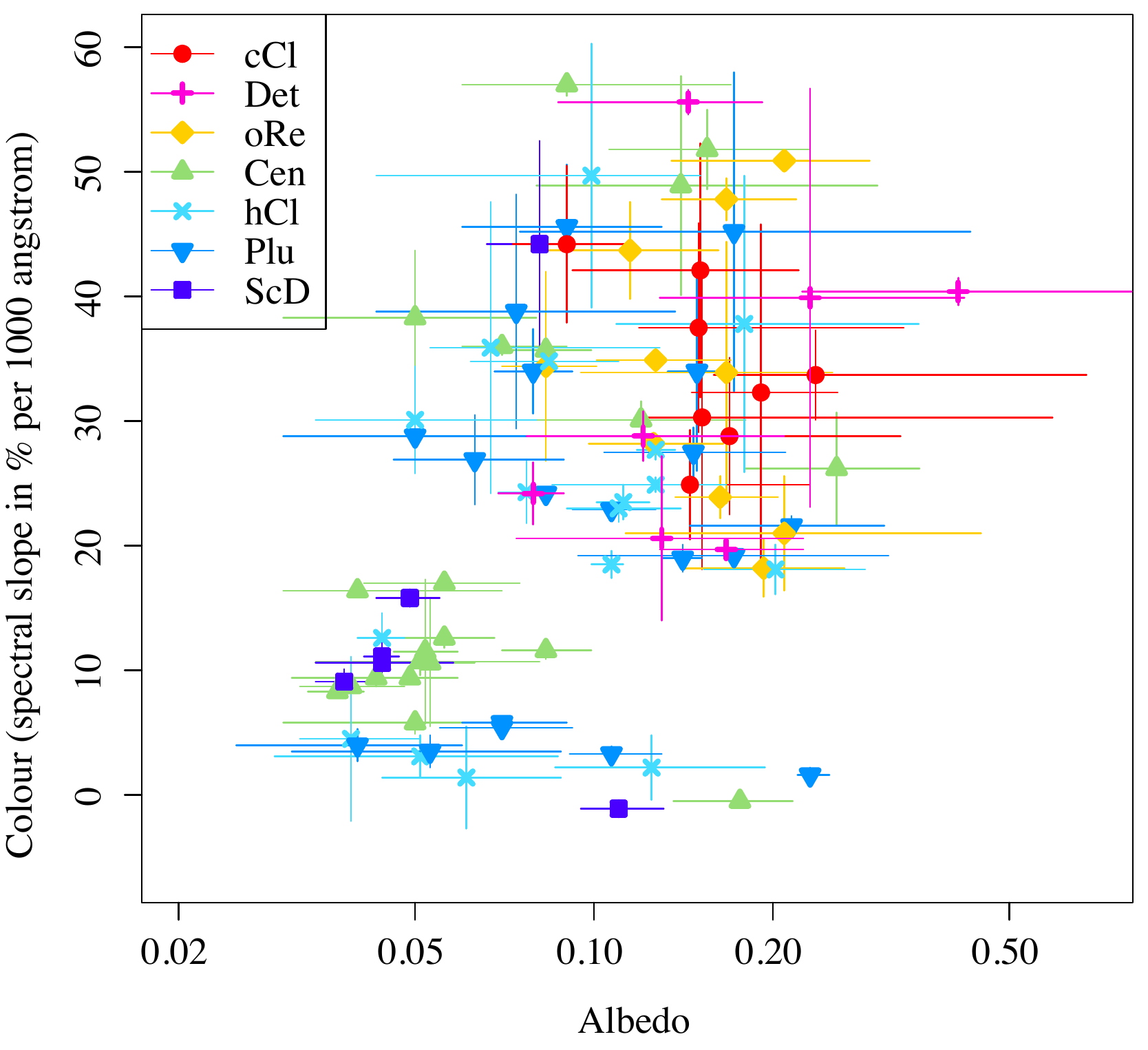}
\caption{Albedo and colour distribution of intermediate-sized KBOs
    in the Herschel sample. Selected dynamical classes are plotted:
    Cen=Centaurs, Plu=Plutinos, ScD=Scattered Disc, hCl=Hot
    Classicals, Det=Detached, cCl=Cold Classicals, oRe=Outer Resonants
  (beyond the classical belt). For a description of these dynamical
classes see \citet{2008ssbn.book...43Gladman}.}
  \label{Fig.AS}
\end{centering}
\end{figure}

KBOs possess the most diverse surfaces of any small body population in
the solar system \citep{1996AJ....112.2310Luu,
1998AJ....115.1667Jewitt}. This is puzzling, given the low and slowly
varying temperatures in the Kuiper belt region. The migration-driven
instability event described above offers a plausible solution to this
problem by allowing planetesimals formed in the \mbox{$\sim$10 to 30 AU}
region to be scattered to the current Kuiper belt. However, the
stochastic character of this process complicates the mapping of
current KBO surface properties to their formation location. A recent
survey of KBOs and Centaurs using {\em Herschel}
\citep{2009EM&P..105..209Mueller} has built on an earlier {\em
Spitzer} survey \citep{2008prpl.conf..161Stansberry} to produce the
largest sample of albedos and sizes for this type of object (Table
\ref{Table.One}). These data reveal that KBOs are also extremely
diverse in reflectivity, with albedos spanning more than an order of
magnitude from about 0.03 to 0.40. An interesting trend emerges when
combining the KBO colour and albedo data \citep{2041-8205-793-1-L2}:
an albedo-colour plot (Figure \ref{Fig.AS}) shows that KBOs fall into
two main groups: a broad bright-red (BR) cluster, centred around
albedo $\sim$0.15 and colour spectral slope $\sim$30\% and a more
compact dark-neutral (DN) clump near albedo $\sim$0.05 and colour
slope $\sim$10\%.  Importantly, KBOs in dynamical classes believed to
originate beyond Neptune occupy only the bright-red group, while KBOs
that may have been scattered out to the Kuiper belt from nearer the
Sun are found in both the bright-red and the dark-neutral groups.
This result hints at a compositional gradient that was present in the
planetesimal disc prior to the planetary instability event.

\begin{table}
 \caption{\label{Table.One}Summary of the KBO/Centaur Herschel Sample Properties. Columns are 1) Dynamical class \citep{2008ssbn.book...43Gladman}, 2) median albedo and central 68\% interval, 3) mean spectral slope in \%/(1000~\AA) and standard deviation. Statistics includes measurement uncertainties by bootstrap resampling and excludes dwarf planets and Haumea-type KBOs. 4) Dominant surface types (BR=bright-red, DN=dark-neutral) present in class (see caption for Figure~\ref{Fig.AS})}
 \begin{indented}
\item[]\begin{tabular}{@{}lllc}
\br
 Dynamical class & Albedo & Colour & Surface Types\\
  \mr
  Scattered Disc   & $0.05^{+0.04}_{-0.01}$ & $16.3\pm12.6$ & DN, BR \\
  Centaurs         & $0.06^{+0.07}_{-0.02}$ & $21.5\pm16.5$ & DN, BR \\
  Hot Classicals   & $0.08^{+0.05}_{-0.04}$ & $22.8\pm15.6$ & DN, BR \\
  Plutinos         & $0.09^{+0.07}_{-0.04}$ & $20.1\pm15.4$ & DN, BR \\
  Outer Resonants  & $0.13^{+0.09}_{-0.05}$ & $31.6\pm12.8$ & BR \\
  Cold Classicals  & $0.15^{+0.08}_{-0.06}$ & $33.2\pm10.3$ & BR \\
  Detached KBOs    & $0.17^{+0.20}_{-0.09}$ & $33.2\pm14.6$ & BR \\
  \br
\end{tabular}
\end{indented}
\end{table}

\section{Long-Term Evolution}
Planetary system development is often considered in the context of {\it dynamical stability}, or the ability of a planet, moon, or asteroid to adhere to its regular motion -- i.e. approximately retain its original orbit -- over time.  Past or future instability can help constrain and explain observations and place our solar system today into context \citep{2013arXiv1311.6816D}.  In this section, we ask, are planetary systems - and in particular the solar system - stable
as their host stars evolve?

\subsection{Long-Term Evolution of the Solar System}

The future orbital evolution of the solar system, like all  other planetary systems, cannot be determined to infinite precision. Instead, we obtain likelihoods for particular futures through suites of numerical simulations.  Qualitatively, the evolutionary behaviour can be split into two parts depending on the state of the Sun: (i) the main sequence and (ii) post-main-sequence phases of Solar evolution.


The Sun will exist in its present state for about a further 6 Gyr  before turning off the main sequence.  During these 6 Gyr, the eight planets will technically evolve chaotically \citep{suswis1988,laskar1989} but realistically are very likely to avoid dynamical instability. In fact, through a suite of simulations, \cite{laskar2008} determined that the inner four planets have a 98\%-99\% chance of surviving and not having planetary orbits cross; 100\% of all his simulations yielded survival of the outer four planets.  Figure 6 shows how the eccentricity of the Earth and Mercury evolve over this timescale. Earth's eccentricity remains nearly constant. Mercury's eccentricity gradually increases, but rarely becomes large enough for its orbit to cross that of Venus.

\begin{figure}
\begin{centering}
  \includegraphics[width=8cm]{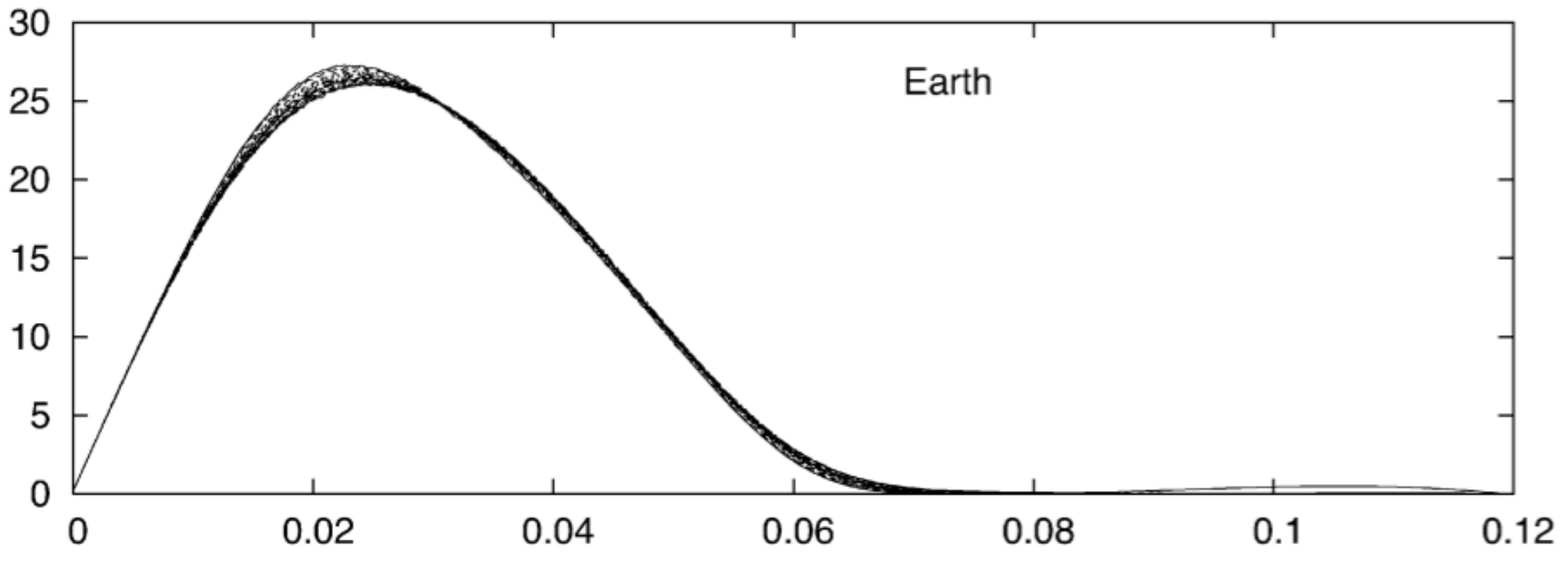}
 \includegraphics[width=8cm]{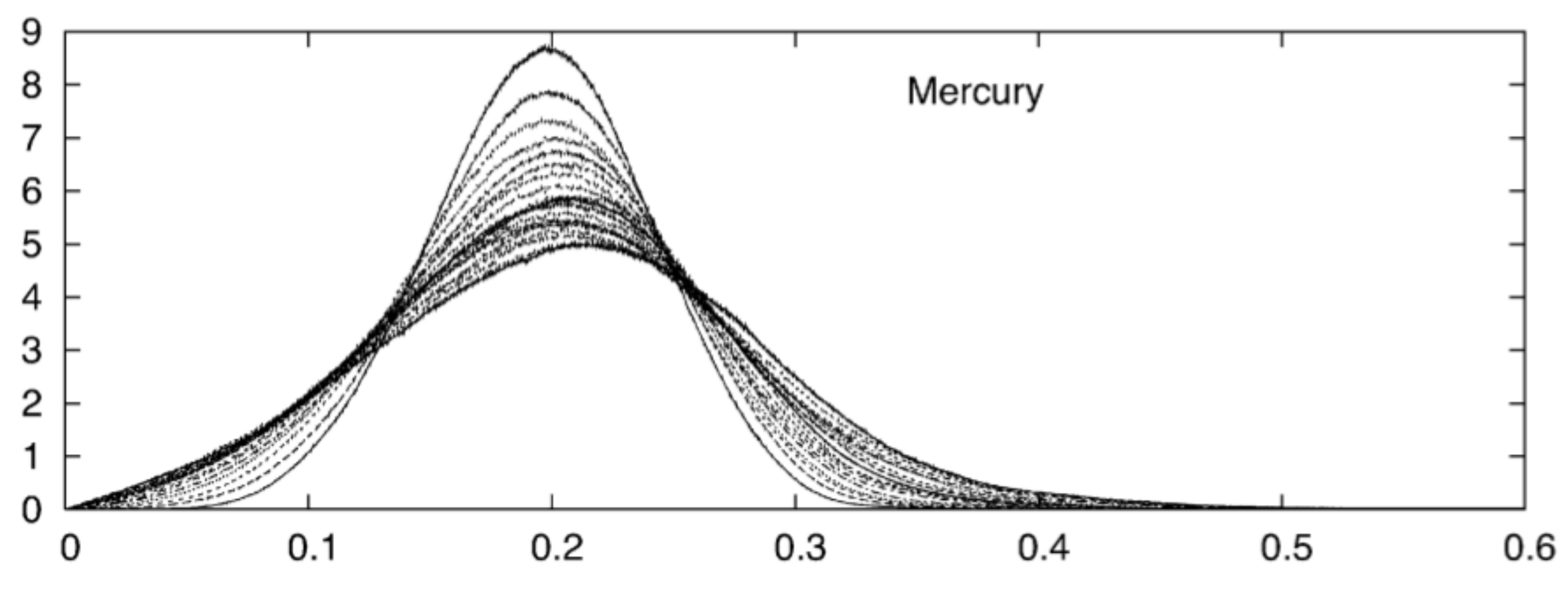}

\caption{
From Figure 9 of Laskar (2008):  Normalised density functions for the eccentricities of Earth and Mercury. On each plot the 19 curves represent intervals of 250 Myr. Each curve is based on 1001 solutions with very close initial conditions. The variation of these curves reflects the chaotic diffusion of the solutions.  Mercury's eccentricity is clearly shown on average to increase with time, whereas the Earth's eccentricity remains relatively constant.}
\label{lasFig}
\end{centering}
\end{figure}


After the Sun leaves the main sequence, it will simultaneously shed about half of its mass while becoming a red giant star that is nearly 1 AU in radius. The increase in radius can more easily allow the Sun to tidally capture objects encountered by the Solar envelope \citep{viletal2014}, whereas the mass loss will push out and potentially warp the orbits of surviving bodies \citep{hadjidemetriou1963}.

The result on the inner three planets will be dramatic, and on the outer five less so.  Mercury and Venus will be swallowed and Earth will suffer an uncertain fate.  Our home planet will lie on the cusp of being tidally drawn into the Sun \citep{schcon2008}.  Mars's orbit will simply expand.  All 4 outer planets will also survive and expand their orbits such that the ratios of their mutual separations will remain unchanged \citep{dunlis1998}.  Beyond Neptune and the Kuiper belt, the Sun's mass loss will trigger instability in the Oort cloud, allowing for objects beyond $10^3 - 10^4$ AU to escape the solar system \citep{verwyat2012}.

\subsection{Long-Term Evolution of Other Planetary Systems}

Observational data for extrasolar planets are significantly less accurate and precise than those for the solar system planets \citep{perryman2011}.  Consequently, the future of exoplanetary systems is not as well constrained.  However, every exoplanetary system so far discovered contains fewer planets than the solar system, simplifying the analysis.


The majority of known planetary systems contain either two or three planets (www.exoplanets.org), motivating study of these two cases in detail. During the main sequence evolution of the host star two planets will never cross orbits if they are sufficiently far away from each other to be {\it Hill stable} \footnote{The concepts of Hill and Lagrange stability are applicable to 3-body systems (2-planet systems) only.  In a Hill stable system, the orbits of both planets can never cross.  A Lagrange stable system is Hill stable and also does not allow the outer planet to escape the system nor the inner planet to crash into the star.} \citep{gladman1993}.  A Hill stable system could potentially become {\it Lagrange unstable} if the inner planet collides with the star or the outer planet escapes the system \citep{bargre2006}.  Empirical and analytical estimates for these stability boundaries, even when scaled down in mass to the test particle limit, show good agreement with numerical simulations of the long-term evolution \citep{wisdom1980,muswya2012,decetal2013,giuetal2013,vermus2013}.

Three-planet systems do not appear to admit analytical stability boundaries, and therefore investigations of these systems require a numerical approach.  These studies have yielded portraits of instability times as a function of initial separation, as well as empirical estimates for these times \citep{chaetal1996,marwei2002,chaetal2008}. These instabilities can occur any time during the main sequence evolution, and fundamentally change the resulting dynamical architecture, perhaps leading to the establishment of a population of hot Jupiters and eccentric planets \citep{beanes2012}.  The study of systems with a higher number of exoplanets produce a similarly-wide variety of outcomes to the three-planet case \citep{smilis2009}.


\begin{figure}
\begin{centering}
\includegraphics[width=8cm]{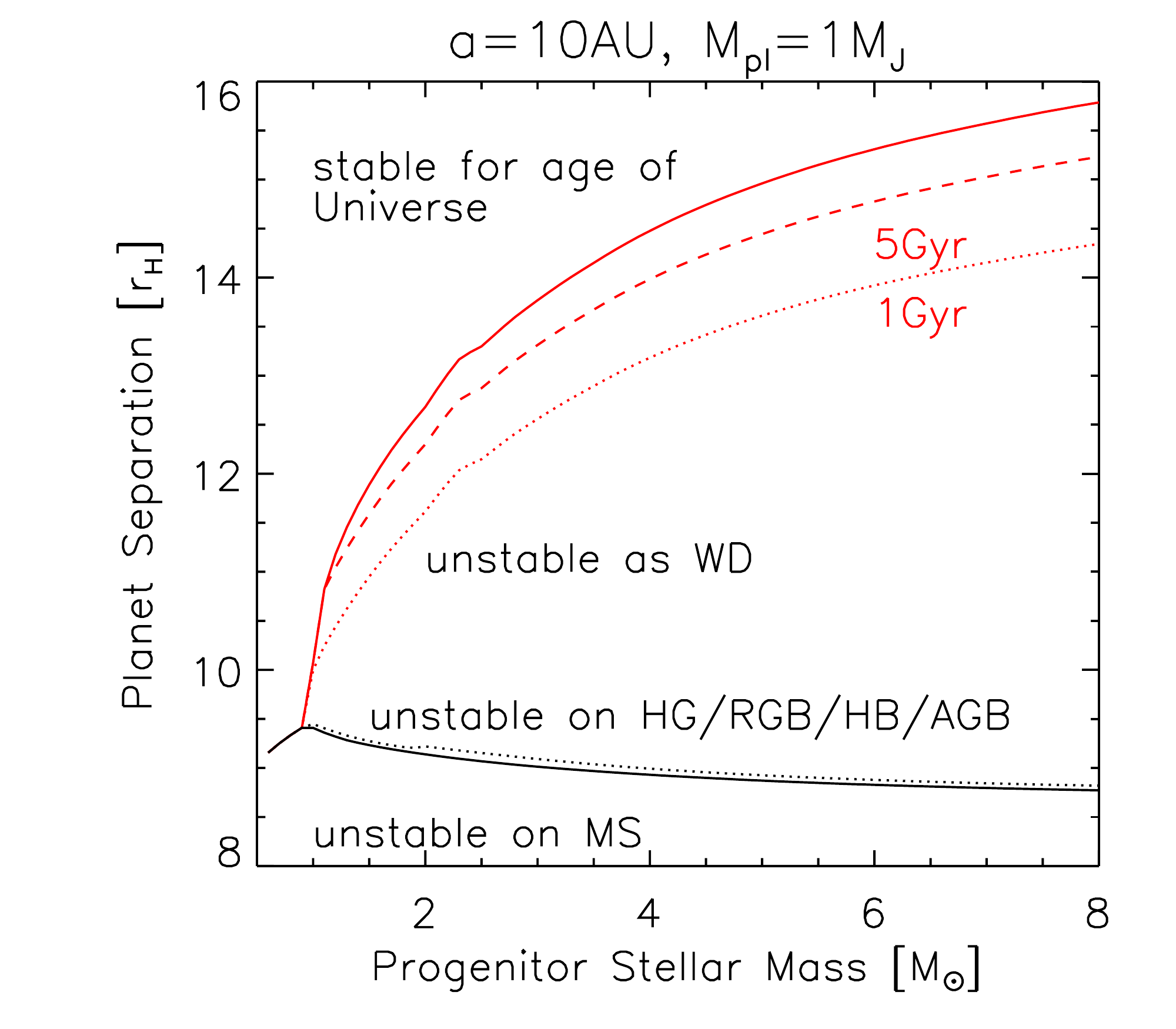}
\caption{
Fig. 1 of Mustill, Veras \& Villaver (2014):  
Approximate stability behaviour of three-planet systems as a function of stellar mass and separation on the main sequence (MS). The quantity $r_H$ on the y-axis represents the single-planet Hill radius.  The planet mass and the innermost semi-major axis are fixed at 1 $M_J$ and 10 AU, respectively. Systems below the black solid line will be unstable on the MS. Systems in the tiny strip between the black solid and the black dotted lines will become unstable at some point between the end of the MS and the end of the asymptotic giant branch (AGB). Systems below the red solid line will be unstable during the WD phase, assuming that the star formed at the big bang; systems above this line must be stable for the age of the Universe. The red dotted and dashed lines show the stability boundaries for WD cooling ages of 1 and 5 Gyr. This plot demonstrates that after the star has become a white dwarf, many more systems can become unstable.}
\label{musFig}
\end{centering}
\end{figure}

We know that planetary systems do survive the giant branch phases of stellar evolution based on abundant evidence of atmospheric metal pollution in white dwarfs (WDs) \citep{zucetal2010,koeetal2014};  WDs, which are the size of the Earth but with the mass of the Sun, represent the end-product of stellar evolution.  In over 35 systems, WD atmospheric metal pollution is accompanied by compact accretion discs of gas and/or dust \citep{gaeetal2006,faretal2009}. Although the precise origin of the pollution and discs have yet to be identified, they must arise from currently dynamically active planetary system remnants.

During giant branch mass loss, the dynamical stability limits of a multi-body system are changed \citep{debsig2002}.  Consequently, two-planet systems, which remained stable throughout the entire main sequence may become unstable either during the giant branch phase, or the WD phase \citep{veretal2013}.  For three-planet systems, the situation becomes more complex, allowing for multiple instabilities to occur during different phases of stellar evolution \citep{musetal2014}.  See Figure 7 for a phase portrait of stability limits through all phases of stellar evolution for three-planet systems.  Planets residing close to the giant star will tidally interact in a complex way depending on the pulsations of the giant \citep{musvil2012}.

A likely progenitor of the WD discs are exo-asteroids.  Although unobservable during main sequence and giant branch phases, their evolution through these epochs will crucially determine their placement in WD systems.  The effect of mass loss on a single planet and single exo-asteroid belt has been considered in the context of an exo-Kuiper belt at about 30 AU \citep{bonetal2011} and an exo-asteroid belt at about 5 AU \citep{debetal2012,frehan2014}.

\section{Summary}

Combining the evidence from these different sources for the early history of the solar system the following picture emerges:

The geochemical and isotopic inventory of meteorites provides a unique chronological record of early solar system evolution. It took less than one million years after condensation of the first solid matter to form the first generation of planetesimals. These planetesimals underwent internal heating between 1 and 4 million years after the solar system formation, largely caused by $^{26}$Al decay, triggering their differentiation into metal core and silicate mantle. The parent asteroids of undifferentiated chondritic meteorites formed 1-2 million years later, and therefore these bodies did not differentiate. Formation of the larger terrestrial planets in the inner solar system likely occurred via collisions of smaller asteroidal bodies. The timescales involved in planetary growth are of the order tens of millions of years, with Mars being a possible exception due to gravitational interaction with nearby Jupiter. There is growing evidence that the volatile element inventory of the inner solar system was added later, i.e. during the last stages of planetary growth.

The presence of the remnants of short-lived radioactive nuclei in meteorites was for a long time interpreted as evidence for the formation of the Sun in a large star cluster. It was usually assumed that the different radioactive nuclei were formed in one single event in a supernova explosion. However, there now exist arguments to support that a sequence of events took place. Here the SLRs with the longest half-lives originate from star complexes, while $^{60}$Fe ($T_{1/2}$ = 2.6 Myr) originates from a molecular cloud and $^{26}$Al ($T_{1/2}$ = 0.72 Myr) from the wind of a local massive star. In this picture the distribution of SLRs in the solar system is the result of sequential star formation in the ISM.

The observation of young disc-surrounded stars also imposes constraints on the timescale of planet formation. The disc fractions in star clusters strongly indicate that most stars dissipate their discs within 2-3 Myr which would favour rapid planet formation - at least for the gas giants.
However, recent works argue in favour of longer disc dissipation times, of the
order of 5-10 Myr. Here a comparison with the meteorite record could possibly bring clarification in the future.

Comparing the typical masses of the  discs around young stars it is found a large fraction of young solar analogues are hosting a disc with a mass well above the minimum mass solar nebula and in principle are capable of forming a planetary system similar to our own. For the most massive of these discs the observed surface densities distribution of the disc mass is roughly consistent with the average surface density of our own solar system. However, most systems are larger than our own solar system, resulting in a lower surface density. It is currently unclear whether this
difference is real or due to the large uncertainties in the observations.

The formation of planets from dust and ice particles likely proceeds in the initial stages by sticking collision, the so-formed porose aggregates eventually compactify. The continued growth from mm-sized particles faces the the formidable bouncing and radial drift barrier. Planetesimals form despite these difficult circumstances, likely by a combination of particle sticking and gravitational collapse of regions overdense in pebble-sized particles.
The various mechanisms which allow particles to avoid this process have been suggested. 
Future studies of the properties of the remnant planetesimals in the asteroid belt and in the Kuiper belt will be crucial to shed light on this open issue.  In addition, studying protoplanetary discs will  help to understand this phase of planet formation better, as populations of mm-cm sized pebbles have been detected in discs around young stars.

It is presently unclear to what degree nearby others stars influence the formation of planetary systems.   From the SLR abundances and the size of the solar system it can be concluded that the birth cluster of the Sun contained at least 2000 stars. Measuring detailed abundance patterns
several groups have recently started the search for other stars that originate from the same gas cloud as the Sun. There are currently some candidate siblings, however further studies are required to confirm those results.

The orbital structure of the Kuiper belt and the cratering record on
the Moon are evidence for a dynamically active early solar system.
Planetary migration and resonance-driven instabilities have acted on
the planetesimal disc to enhance impact rates onto the inner solar
system, to boost the mixing of populations formed at different
distances from the Sun, and to sculpt the dynamics of the main
planetesimal reservoirs. On the other hand, external forces from close stellar encounters within the Sun´s parental cluster could have been the cause of episodic dynamical excitations of minor body populations and impact rates. The current challenge for understanding the
origin of the solar system (and other planetary systems) is to seek
clues that allow us to see past these dynamical events, to the time
when the first planetesimals began to form.

What will happen to the solar system in the future? After the Sun leaves the main sequence Mercury and Venus will be swallowed and Earth will suffer an uncertain fate as it  will lie on the cusp of being tidally drawn into the Sun.  Mars and the 4 outer planets will also survive and expand their orbits.

In summary, in recent years we have seen many new insights into the formation of the solar system. However, there are still a considerable number of open questions where different competing theories exist. 
So far the different fields contributing in the investigations to the origin of the solar system have generally worked fairly orthogonally. However, it would be timely to start combining these seperate efforts, using them to build a more complete pictures. Progress can be expected in determining the actual chronology of events.

\section*{Acknowledgements} 
We want to thank both referees for their very constructive comments. SP is supported by the Minvera program of the Max-Planck society. 
MBD is supported by the Swedish Research Council (grant 2011-3991). AJ was supported by the Knut and Alice Wallenberg Foundation, the Swedish Research Council (grant 2010-3710) and the European Research Council (ERC Starting Grant 278675). LT is partly supported by the Italian Ministero dell’Istruzione, Universita` e Ricerca through the grant iALMA - Progetti Premiali 2012  (CUP C52I13000140001). MT acknowledges support by Deutsche Forschungsgemeinschaft (DFG), and Klaus Tschira Stiftung gGmbH. DV is supported from the European Research Council under the European Union’s Seventh Framework Programme (FP/2007-2013) / ERC Grant Agreement n. 320964 (WD
Tracer).

\bibliographystyle{harvard}

\end{document}